\documentstyle[prl,aps,epsfig,multicol,amssymb]{revtex}
\begin{document}
\draft
\begin{multicols}{2}
{\bf \noindent Comment on ``Phase Diagram of an Asymmetric Spin Ladder.''
} \vspace{10pt}

In a recent letter S. Chen {\em et al.} \cite{voit} investigated
the so-called asymmetric spin ladder, a spin-half
Heisenberg chain with alternation in the next-nearest-neighbor 
(n.n.n.) interaction \cite{note}.
Based on bosonization and renormalization group analysis, 
they claimed that in the limit
of small frustration ($J_2/J_1$) the asymmetry in the n.n.n. integrals
destabilizes the isotropic Heisenberg fixed point leading to
a new phase with gapless excitations and vanishing spin-wave
velocity.

In this comment, using Bethe {\em ansatz}  and conformal field theory
we show instead  that the n.n.n. spin-Peierls operator 
\begin{equation}\label{nnn}
{\hat O}_{\rm n.n.n.}=\sum_{l} (-1)^l {\hat {\bf S}}_l\cdot{\hat{\bf S}}_{l+2},
\end{equation}
represents an {\em irrelevant} perturbation for the Heisenberg
chain in the regime of weak frustration. Since the latter operator
is the one associated with the alternation in the n.n.n. exchange,
 this clearly invalidates the mentioned  
claim and the conclusions of their paper.

\begin{figure}
\begin{center}
\epsfig{figure=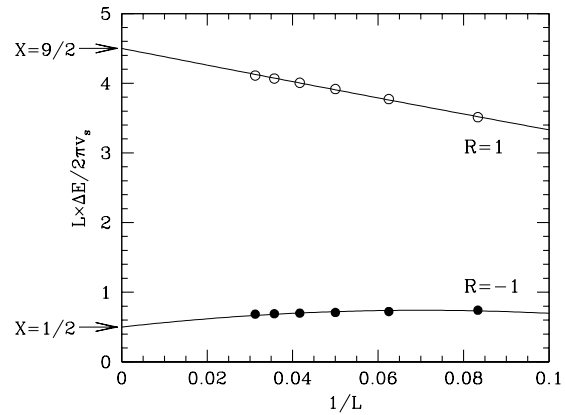,width=8cm}
\begin{minipage}{8cm}
\vspace{-0.6 cm}
\caption{\label{betofig} 
Size scaling of the gap of the lowest eigenvalues of the 
isotropic spin-half Heisenberg
chain ($J_2/J_1=0$) with $S^z=0$, $k=\pi$, even parity under spin reflections,
and $R=\pm 1$ under lattice reflections. Lines are guides for the eye.}
\end{minipage}
\end{center}
\end{figure}

For sake of generality, we refer to the anisotropic XXZ chain.
In order to study the relevance of the operator (\ref{nnn}) we have to
identify the quantum numbers $\{j\}$ (referenced to the ground state with
energy $E_0$) of the intermediate states appearing
in the Lehmann representation of the associated susceptibility.
These are total spin $S^z=0$, momentum $k=\pi$, even parity under 
spatial reflection  ($l\to -l$) $R=1$, and even parity under 
spin-reflection [$(S^x,S^y,S^z)\to (-S^x,S^y,-S^z)$]. 
The only difference with the {\em nearest-neighbor} spin-Peierls operator,
a well-known relevant perturbation,
is the spatial-reflection quantum number, $R=-1$.
As it is known by conformal field theory, the scaling dimension  $X$
of a given operator
(where $X<1$ characterizes a relevant operator) is related
to the finite-size corrections of the energy of the lowest intermediate
eigenstate $j$ by the relation \cite{bibbia} 
\begin{equation}
\Delta E(L)=E_{j}(L)-E_0(L) = 2\pi v_s X/L~,
\end{equation}
where 
$v_s$ is the spin-wave velocity and $L$ is the number of sites 
of the ring.
The finite-size corrections can be computed either by Bethe {\em ansatz}
(for $J_2/J_1=0$) or by bosonization in the Luttinger regime ($J_2/J_1\lesssim
0.241$). 
For the nearest-neighbor spin-Peierls operator, due to the existence
of a low-lying eigenstate with the correct quantum  numbers, this procedure gives
$X=K$, where $1/2\leq K\leq 1$ (with $K=1/2$ at the isotropic point) 
is the dimensionless coupling constant of the model.
In contrast, the finite-size spectrum of the Heisenberg model
does not contain low-lying states simultaneously even under both
spatial- and spin-reflection, yielding $X=9K>1$, showing the
irrelevance of the operator (\ref{nnn}).
This is illustrated in Fig.~\ref{betofig} using
Lanczos exact diagonalization technique. However we stress that 
our conclusions follow directly from the exact analytical 
solution of the spin-half Heisenberg chain.

The recent bosonization analysis of Sarkar and Sen \cite{cinesini}
is consistent with our conclusion.

\vspace{-2.cm}

This work was partially supported by MURST (COFIN01).
L.C. was supported by NSF grant DMR-9817242.
A.P., L.C., and F.B. would like to acknowledge 
kind hospitality at SISSA.

\vspace{0.2cm} 

\noindent Luca Capriotti,

        Kavli Institute for Theoretical Physics,

        University of California,

        Santa Barbara, CA 93106-4030

\vspace{0.2cm}

\noindent Federico Becca,

        Institut de Physique Th\'eorique, 

        Universit\'e de Lausanne,

        CH-1015 Lausanne, Switzerland

\vspace{0.2cm} 
\noindent Sandro Sorella,

        Istituto Nazionale di Fisica della Materia (INFM), and

        International School for Advanced Studies (SISSA),

        Via Beirut 4, I-34014 Trieste, Italy

\vspace{0.2cm} 

\noindent Alberto Parola,

        Istituto Nazionale di Fisica della Materia (INFM), and

        Universit\`a dell'Insubria, via Valleggio 11,
  
        I-22100 Como, Italy

\vspace{0.2cm} 


\end{multicols}


\vspace*{-0.5cm}
\begin{thebibliography}{99}
\vspace*{-1cm}
\bibitem{voit} S. Chen, H. B\"uttner, and J. Voit, \prl {\bf 87}, 
087205 (2001).
\bibitem{note} According to the standard notation  $J_1$ and $J_2$ indicate
the nearest and next-nearest-neighbor antiferromagnetic couplings, respectively.
\bibitem{bibbia} J.L. Cardy, Nucl. Phys. {\bf B270} [FS16], 186 (1986).
\bibitem{cinesini} S. Sarkar, and D. Sen, \prb {\bf 65} 172408 (2002).
\end{thebibliography}
\end{document}